\documentclass[twoside,leqno,twocolumn]{article}  
\usepackage{ltexpprt} 
\usepackage{algpseudocode}
\usepackage{algorithm}
\usepackage{amsmath}
\usepackage{subfig}
\usepackage{graphicx}
\usepackage{color}
\usepackage[svgnames]{xcolor}

\usepackage{amssymb}

\begin{document}
\newcommand\blfootnote[1]{%
  \begingroup
  \renewcommand\thefootnote{}\footnote{#1}%
  \addtocounter{footnote}{-1}%
  \endgroup
}
\renewcommand*{\thefootnote}{\fnsymbol{footnote}}

\title{\Large Scalable Positional Analysis for Studying Evolution of Nodes in Networks}
\author{Pratik Vinay Gupte\thanks{\scriptsize Indix India, IIT Madras Research Park, Chennai, India 600 113. Email: pratik.gupte@gmail.com. This work was done by the author when he was a Research Scholar at the Department of Computer Science and Engineering, IIT Madras.}
\and 
Balaraman Ravindran\thanks{\scriptsize Department of Computer Science and Engineering, Indian Institute of Technology Madras, Chennai, India 600 036. Email: ravi@cse.iitm.ac.in.}} 
\date{}

\maketitle


\begin{abstract} \small\baselineskip=9pt In social network analysis, the fundamental idea behind the notion of \emph{position and role} is to discover actors who have similar structural signatures. Positional analysis of social networks involves partitioning the actors into disjoint sets using a notion of equivalence which captures the structure of relationships among actors. Classical approaches to Positional Analysis, such as Regular equivalence and Equitable Partitions, are too strict in grouping actors and often lead to trivial partitioning of actors in real world networks. An $\epsilon$-Equitable Partition ($\epsilon$EP) of a graph, which is similar in spirit to Stochastic Blockmodels, is a useful relaxation to the notion of structural equivalence which results in meaningful partitioning of actors. In this paper we propose and implement a \emph{new scalable distributed} algorithm based on MapReduce methodology to find $\epsilon$EP of a graph. Empirical studies on random power-law graphs show that our algorithm is highly scalable for sparse graphs, thereby giving us the ability to study positional analysis on very large scale networks. We also present the results of our algorithm on time evolving snapshots of the \emph{facebook} and \emph{flickr} social graphs. Results show the importance of positional analysis on large dynamic networks. 
\end{abstract} 

\blfootnote{\scriptsize The authors were partly funded by a grant from Ericsson Global Research. This manuscript is the pre-final version of the work which has been accepted at the workshop on Mining Networks and Graphs: A Big Data Analytic Challenge, to be held in conjunction with the SIAM Data Mining Conference in April 2014 (SDM-14).}
\section{Introduction}
Positional Analysis (PA) \cite{wasserman1994sna,borgatti1992notions} has been long used by sociologists to draw \textit{equivalence classes} from network of social relationships. Finding equivalences in the underlying social relations gives us the ability to model social behaviour, which further aids drawing out the social and organizational structure prevalent in the network. In PA, actors who have same structural correspondence to other actors in a network are said to occupy same ``position''. As an example, head coaches in different football teams occupy the position \textit{manager} by the virtue of the similar kind of relationship with players, assistant coaches, medical staff and the team management. It might happen that an individual coach at the position \textit{manager} may or may not have interaction with other coaches at the same position. Given an organizational setting and the interaction patterns that exist amongst the individuals of this organization, we naturally tend to draw some abstraction around the structure and try to model its behaviour. For example, in our football team setting, the actors at the position \textit{manager} can be a ``Coach'' to actors at the position \textit{player} or  a ``Colleague'' to the actors at the position \textit{assistant coach}. Similarly, the actors at position \textit{medical staff} can be a ``Physiotherapist'' or a ``Doctor'' to actors at the position \textit{player}. While positional analysis is a very intuitive way of understanding interactions in networks, this hasn't been widely studied for large networks due to the difficulty in developing tractable algorithms. In this paper we present a positional analysis approach that has good scaling behaviour. Hence, this work opens up the study of positions in large social networks.

The key element in finding positions, which aid in the meaningful interpretation of the data is the notion of \textit{equivalence} used to partition the network. Classical methods of finding equivalence like structural equivalence \cite{lorrain1971}, regular equivalence \cite{whitereitz1983}, automorphisms \cite{everett1985role} and equitable partition \cite{mckay1981} often lead to trivial partitioning of the actors in the network. This trivial partitioning of actors is primarily attributed due to either their strictness in the case of structural, automorphisms and equitable equivalence, which results in largely singleton position; or their leniency in the number of connections the actors at each position can possibly have with the actors at another position, as in the case of regular equivalence, which results in a giant equivalence class.

An $\epsilon$-equitable partition ($\epsilon$EP) \cite{kate2009comad} is a notion of equivalence, which has many advantages over the classical methods. $\epsilon$EP allows a leeway of $\epsilon$ in the number of connections the actors at a same position can have with the actors at another position. The notion of $\epsilon$EP is similar in spirit to the notion of stochastic blockmodels \cite{stochastic1987}, in that both approaches permit a bounded deviation from perfect equivalence among actors. In the Indian movies dataset from IMDB, authors in \cite{kate2009comad} have shown that actors who fall in the same cell of the partition, tend to have acted in similar kinds of movies. Further, the authors also show that people who belong to a same position of an $\epsilon$EP tend to evolve similarly. In large social networks, tagging people who belong to the same position has potentially many advantages, both from business and individual perspective, such as, position based targeted promotions, ability to find anomalies, user churn prediction and personalised recommendations.

Though efficient graph partition refinement techniques and their application in finding the regular equivalence of a graph are well studied in the graph theoretic literature \cite{cardon1982, paige1987three}, the application of these techniques for doing positional analysis of very large social graphs and networks is so far unknown. In this work, we propose a new algorithm to find the $\epsilon$-equitable partition of a graph and focus on scaling this algorithm. We have successfully validated our algorithm with detailed studies on \textit{facebook} social graph, highlighting the advantages of doing positional analysis on time evolving social network graphs. We present few results on a relatively large component of the \textit{flickr} social network. Further more, the empirical scalability analysis of the proposed new algorithm shows that the algorithm is highly scalable for very large sparse graphs. 

The contribution of our work has been twofold. First, we propose a new algorithm with better heuristics for finding the $\epsilon$-equitable partition of a graph. Second, we have been able to parallelize this algorithm using MapReduce \cite{dean2008mapreduce} methodology, which gives us the ability to study Positional Analysis on large dynamic social networks.

The rest of the paper is organized as follows. Section \ref{lbl:maths} discusses few mathematical preliminaries along with the definition of $\epsilon$EP. We discuss the Fast and Parallel $\epsilon$EP Algorithm and implementation in Section \ref{lbl:feep}. We present the scalability analysis, evaluation methodology, dataset details and experimental results in Section \ref{lbl:results}. Section \ref{lbl:conclusion} concludes the paper with possible future directions. 

\vspace{-8pt}
\section{Mathematical Preliminaries}
\label{lbl:maths}
\begin{Definition}
\textit{(Partition of a graph)}

Given, graph G $\equiv \langle$V, E$\rangle$, V is the vertex set and E is the
edge set.
A partition $\pi$ is defined as $\pi = \{C_1, C_2, ...,C_n\}$ such that,
\vspace{-4pt}
\begin{itemize}
	\item$\cup{C_i} = V, i = 1$ to $n$ and
	\item$ i \neq j \Rightarrow C_i \cap C_j = \phi$
\end{itemize}
\end{Definition}
Thus, the definition of a \textit{partition} of a graph $G$ means that we
have \textit{non-empty} subsets of the vertex set $V$, such that all subset
pairs are \textit{disjoint} to each other. These subsets $C_1, C_2, ...,C_n$ are
called \textit{cells} or \textit{blocks} of the partition $\pi$.

\begin{Definition}
\label{lbl:epdef}
\textit{(Equitable partition)}

A partition $\pi=\{C_1,C_2,...,C_K\}$ on the vertex set $V$ of graph $G$ is said
to be \textit{equitable} \cite{mckay1981} if,
$\text{for all } 1 \leq i,j \leq {K}, 
	{deg_{\textsc{g}}}(u, C_j) = {deg_{\textsc{g}}}(v, C_j), \text{ for all }  u,v \in C_i$,

where, 
\begin{equation}
{deg_{\textsc{g}}}(v_i, C_j) = sizeof\{ v_k \text{ }|\text{ } (v_i, v_k) \in E \text{ and } v_k \in C_j \}
\end{equation}
\end{Definition}
The term ${deg_{\textsc{g}}}(v_i, C_j)$ denotes the number of vertices in cell
$C_j$ adjacent to the vertex $v_i$. 
An equitable partition can be used to define positions in a network; each cell $C_j$ corresponds to a position and ${deg_{\textsc{g}}}(v_i, C_j)$ corresponds to the number of connections the actor $v_i$ has to the position $C_j$.

McKay's algorithm (equitable refinement procedure \cite{mckay1981}) for finding the equitable partition takes as input an ordered partition $\pi$ on $V$ and the graph $G$. The initial partition is usually a \textit{unit} partition, (\textit{i.e.}, all vertices belong to a single cell) of the graph $G$. An \textit{active} list is used to hold the indices of all the unprocessed cells from $\pi$, and is updated in every iteration of the refinement procedure. $c_a$ is the set of vertices from the \textit{current active cell} of the partition $\pi$. The initial active cell $c_a$ of a unit partition is therefore the entire vertex set $V$. Additionally, a function $f$, which maps every vertex $u \in V$ to its degree to $c_a$ is used. Mathematically, $f: V \rightarrow \mathcal{N}$ is defined as
follows:
\begin{equation}
\label{eq:fou}
\begin{split}
	f&(u)= {deg_{\textsc{g}}}(u,c_a)\text{ }\forall u\in V
\end{split}
\end{equation}
The procedure then sorts (in ascending order) the vertices in each cell of the partition using the value assigned to each vertex by the function $f$ as a key for comparison. The procedure then \textit{splits} the
contents of each cell wherever the keys differ, thereby creating new cells. The partition $\pi$ is updated accordingly, and the indices corresponding to any new cells formed after the split are added to the \textit{active} list. The procedure exits when the \textit{active} list is empty. The resulting partition is the \textit{coarsest equitable partition} of the graph $G$.

\begin{Definition}
\label{eq:eep}
\textit{($\epsilon$-Equitable partition)}
A partition $\pi_\epsilon = \{C_1, C_2, ...,C_{K}\}$ of the vertex set \{$v_1,v_2,...,v_n$\}, is defined as $\epsilon$-\textit{equitable partition} if, $\text{for all }1 \leq i,j \leq {K}, \arrowvert deg_{\textsc{g}}(u, C_j) - deg_{\textsc{g}}(v, C_j) \arrowvert \text{ }\leq \text{ }\epsilon, \text{ for all }u,v \in C_i$,
\end{Definition}
The above definition proposes a relaxation to the strict partitioning condition of equitable partition (Definition \ref{lbl:epdef}), now equivalent actors can have a difference of $\epsilon$ in the number of connections to other cells in the partition.

\begin{Definition}
\textit{(Degree vector of a vertex)}

Given a partition $\pi = \{C_1, C_2, ...,C_{K}\}$ of the vertex set $V$ of a graph $G$, the degree vector of a vertex $u \in V$ is defined as follows,
\begin{multline}
\label{eq:degreevector}
\overrightarrow{deg_{\textsc{g}}}(u) = [{deg_{\textsc{g}}}(u, C_1),{deg_{\textsc{g}}}(u, C_2),...,{deg_{\textsc{g}}}(u, C_K)]
\end{multline}
\end{Definition}
Thus, the \textit{degree vector} of a node $u$ is a vector of size $K$ (the total number of cells in $\pi$), where each component of the vector is the number of neighbours $u$ has in each of the member cells of the partition $\pi$.

\section{Fast and Parallel Epsilon Equitable Partition}
\label{lbl:feep}
\subsection{Problems Addressed}
Kate and Ravindran \cite{kate2009comad} proposed an $O(n^3)$ algorithm to find the $\epsilon$EP of a graph. We discuss this algorithm briefly\footnote{\scriptsize For more details on this algorithm, interested readers are kindly referred to \cite{kate2009comad} Section 3.5.}. Input to this algorithm is $(i)$ the graph, $(ii)$ the \textit{coarsest equitable partition} of graph and $(iii)$ a value of $\epsilon$. The cells in the input equitable partition are arranged by ascending order of their \textit{cell degrees}\footnote{\scriptsize \textit{Cell or block degree} of a cell of an equitable partition is the degree of the member nodes in that cell.}. The algorithm then computes the \textit{degree vector} (Equation \ref{eq:degreevector} for each of the vertices in the graph $G$. The algorithm then tries to merge these cells by taking two consecutive cells at a time. If the degree vectors of the member nodes from these two cells are within $\epsilon$ distance of each other, they are merged into a single new cell. For further merging, this new cell becomes the current cell, which is then compared with the next cell for a possible merger. If the merging fails, the next cell becomes the current cell. The algorithm exits if no further merging of cells is possible. Also, the degree vectors need to be updated whenever two cells are merged. The time complexity of this algorithm to find $\epsilon$EP of a graph is $O(n^3)$.

\subsection{Fast $\epsilon$-Equitable Partition}
The implementation of our Fast $\epsilon$EP algorithm is directly based on the modification of McKay's original algorithm \cite{mckay1981} to find the equitable partition of a graph, which iteratively refines an ordered partition until it is equitable (Section \ref{lbl:maths}, Definition \ref{lbl:epdef}). The key idea in our algorithm is to allow splitting a cell only when the degrees of the member nodes of a cell are more than $\epsilon$ apart. The Fast $\epsilon$EP algorithm and its \textsc{split} function is explained in Algorithm \ref{algo:feep}. The algorithm starts with the \textit{unit} partition of the graph $G$ and the current active cell $c_a$ having the entire vertex set $V$. It then computes the function $f$ (line 5, Algorithm \ref{algo:feep}) for each of the vertices of the graph. The algorithm then calls the \textsc{split} function (line 11, Algorithm \ref{algo:feep}). The \textsc{split} function takes each cell from the partition $\pi$ and sorts the member vertices of these cells using the function $f$ as the comparison key (Equation \ref{eq:fou}, Section \ref{lbl:epdef}). Once a cell is sorted, a linear pass through the member vertices of the cell is done to check if any two consecutive vertices violate the $\epsilon$ criteria. In case of violation of the $\epsilon$ condition, the function \textit{splits} the cell and updates the partition $\pi$ and the active list accordingly. The algorithm exits either when the active list is empty or when $\pi$ becomes a \textit{discrete} partition, \textit{i.e.}, all cells in $\pi$ are singletons. 

\begin{algorithm}[h]
\small
	\caption{Fast $\epsilon$-Equitable Partition}
	\label{algo:feep}
	\textbf{Input:} graph $G$, \textit{ordered unit} partition $\pi$, epsilon $\epsilon$ \\
	\textbf{Output:} $\epsilon$-equitable partition ${\pi}$
	\begin{algorithmic}[1]
		\State active = indices($\pi$)
		\While{(active $\neq$ $\phi$) \textbf{and} ($\pi$ is \textbf{not} $discrete$)}
			\State $idx = min$(active)
			\State active = active $\smallsetminus$ $\{idx\}$
			\State $f(u) = {deg_{\textsc{g}}}(u,$ $\pi[idx]) $ $\forall u\in V$ \Comment $f:$ $V$ $\rightarrow$ $\mathcal{N}$
			\State $\pi{'}$ = \textsc{split}$(\pi,f,\epsilon)$ \Comment \textit{line 11}
			\State active = active $\cup$ [\textit{ordered} indices of newly \textit{split} cells from $\pi{'}$, while replacing (\textit{in place}) the indices from $\pi$ which were \textit{split}]
			\State $\pi = \pi{'}$
		\EndWhile \\
		\Return $\pi$
	\end{algorithmic}
	\hrulefill
	\begin{algorithmic}[1]
	\setcounter{ALG@line}{10}
	\Function{Split}{$\pi,f,\epsilon$}
		\State $idx=0$ \Comment index for return partition ${\pi}_s$
		\For{ \textbf{each} \textit{currentCell} in $\pi$}
			\State $sortedCell$ = \textsc{sort}($currentCell$) using $f$ as the comparison key \Comment \textit{i.e.}, if $f(u) < f(v)$ then $u$ appears before $v$ in $sortedCell$
			\State $currentDegree=f(sortedCell[0])$
			\For{ \textbf{each} $vertex$ in $sortedCell$}
				\If{$(f(vertex) - currentDegree) \leq \epsilon$}
					\State \textit{Add} $vertex \text{ to cell } {\pi}_s[idx]$
				\Else
					\State $currentDegree = f(vertex)$
					\State $idx = idx + 1$
					\State \textit{Add} $vertex \text{ to cell } {\pi}_s[idx]$
				\EndIf
			\EndFor
			\State $idx = idx + 1$
		\EndFor \\
		\Return ${\pi}_s$
	\EndFunction
	\end{algorithmic}
\end{algorithm}
\normalsize

\subsubsection{Time complexity Analysis}
Let $n$ be the size of the vertex set $V$ of $G$. The $while$ loop of line 2 can run at most for $n$ iterations: the case when \textsc{split} leads to the \textit{discrete} partition of $\pi$, hence \textit{active} will have $n$ indices from $[0, 1, ...,(n-1)]$. The computation of the function $f(u) = {deg_{\textsc{g}}}(u,$ $c_a) $
$\forall u\in V$ (line 5, Algorithm \ref{algo:feep}), either takes time proportional to the length of the current active cell $c_a$ or to the length of the adjacency list of the vertex $u$\footnote{\scriptsize Finding the \textit{degree} of a vertex to current active cell translates to finding the \textit{cardinality} of the \textit{intersection set} between the current \textit{active cell} $c_a$ and the \textit{adjacency list} of the vertex $u$. With a good choice of a data structure, the time complexity of intersection of two sets is usually proportional to the cardinality of the smaller set.}. The \textsc{sort} function inside \textsc{split} procedure (line 14, Algorithm \ref{algo:feep}) is bound to $O(nlogn)$. Also, the ``\textit{splitting}'' (line 15 to line 24, Algorithm \ref{algo:feep}) is a linear scan and comparison of vertices in an already sorted list, hence is bound to $O(n)$. Hence, the total running time of the function \textsc{split} is bound to $O(nlogn)$.

The maximum cardinality of the current active cell $c_a$ can at most be $n$. Further, for \textit{dense} undirected simple graph, the maximum cardinality of the adjacency list of any vertex can also at most be $(n-1)$. Therefore for $n$ vertices, line 5 of algorithm \ref{algo:feep} performs in $O(n^2)$. For \textit{sparse} graphs, the cardinality of the entire edge set is of the order of $n$, hence line 5 of algorithm \ref{algo:feep} performs in the order $O(n)$.
 
Therefore, the total running time complexity of the proposed \textit{Fast} $\epsilon$-\textit{Equitable Partitioning} algorithm is $O({n^3})$ for \textit{dense} graphs and $O({n^2}logn)$ for \textit{sparse} graphs. In reality this would be quite less, since subsequent \textit{splits} would only reduce the cardinality of the current \textit{active} cell $c_a$. Which implies that we can safely assume that the cardinality of set $c_a$ will be less than the cardinality of the \textit{adjacency list} of the vertices of the  graph. This analysis is only for the serial algorithm. Empirical scalability analysis on random power-law graphs shows that our parallel algorithm (Section \ref{sec:peep}, Algorithm \ref{algo:peep}) is an order faster in time for sparse graphs.

\subsection{Parallel $\epsilon$-Equitable Partition}
\label{sec:peep}
This section describes the parallel implementation of the Fast $\epsilon$-Equitable Partition Algorithm \ref{algo:feep} by MapReduce methodology \cite{dean2008mapreduce}.

In the Parallel $\epsilon$EP Algorithm, we have implemented the most computationally intensive step of the $\epsilon$EP algorithm, namely, computation of function $f$ (Equation \ref{eq:fou}), as a \textsc{map} operation. Each \textit{mapper} starts by initializing the \textit{current active cell} $c_a$ for the current iteration (\textit{line} 3 Algorithm \ref{algo:peep}). The \textit{key} input to the \textsc{map} phase is the node id $n$ and the node data corresponding to the node $n$ is tagged along as the \textit{value} corresponding to this key. The \textsc{map} operation involves finding the degree of the node $n$ to the current active cell $c_a$, which translates to finding the size of the intersection of the adjacency list of $n$ and the member elements of $c_a$ (\textit{line} 5, Algorithm \ref{algo:peep}). The \textsc{map} phase \textit{emits} the node id $n$ as the \textit{key} and the degree of $n$ to the current active cell $c_a$ as a \textit{value}. This corresponds to the value of function $f$ (Equation \ref{eq:fou}) for the node $n$. Finally, a single \textit{reducer} performs the \textit{split} function (\textit{line 6}, Algorithm \ref{algo:feep}). The output of the \textsc{reduce} phase is used by the $(i)$ \textit{mapper} to initialize the active cell $c_a$ and the $(ii)$  \textit{reducer} itself to update the partition $\pi$ and the \textit{active} list. Single MapReduce step of the algorithm is depicted in Algorithm \ref{algo:peep}. The \textit{iterative} MR job continues till the \textit{active} list becomes empty or the partition becomes \textit{discrete}.

\newcommand{\Statet}[1][1]{\State\hspace{10pt}}
\newcommand{\Statett}[1][1]{\State\hspace{18pt}}
\begin{algorithm}
	\caption{MapReduce step of the Parallel $\epsilon$-Equitable Partition}
	\label{algo:peep}
	\begin{algorithmic}[1]
		\State \textbf{class} \textsc{Mapper}
		\Statet  \textbf{method} \textsc{initialize}()
		\Statett  $c_a\leftarrow$\textit{Current Active Cell}\Comment $active$[0]
		\Statet  \textbf{method} \textsc{map}(id $n$, vertex $N$)
		\Statett  $d\leftarrow|{N.AdjacencyList\cap{c_a}}|$ \Comment $d$ corresponds to the value of function $f(n)$, Equation \ref{eq:fou}
		\Statett  \textsc{emit}(id $n$, value $d$) 
	\end{algorithmic}
	\hrulefill
	\begin{algorithmic}[1]
		\State \textbf{class} \textsc{Reducer}\Comment Single Reducer
		\Statet \textbf{method} \textsc{reduce}()
		\Statett  \textsc{split}($\pi,f,\epsilon$)\Comment Algorithm \ref{algo:feep}, \textit{line 11}
		\Statett  \textsc{update}(\textit{active})\Comment Algorithm \ref{algo:feep}, \textit{line 7}
		\Statett  \textsc{update}($\pi$)
	\end{algorithmic}
\end{algorithm}

\subsubsection{Implementation of the Parallel $\epsilon$EP Algorithm \ref{algo:peep}}
\begin{figure*}[!htpb]
\centering
\vspace{-50pt}
	\subfloat[Conceptual Overview of our Lightweight MapReduce Implementation]{\includegraphics[width=0.9\linewidth]{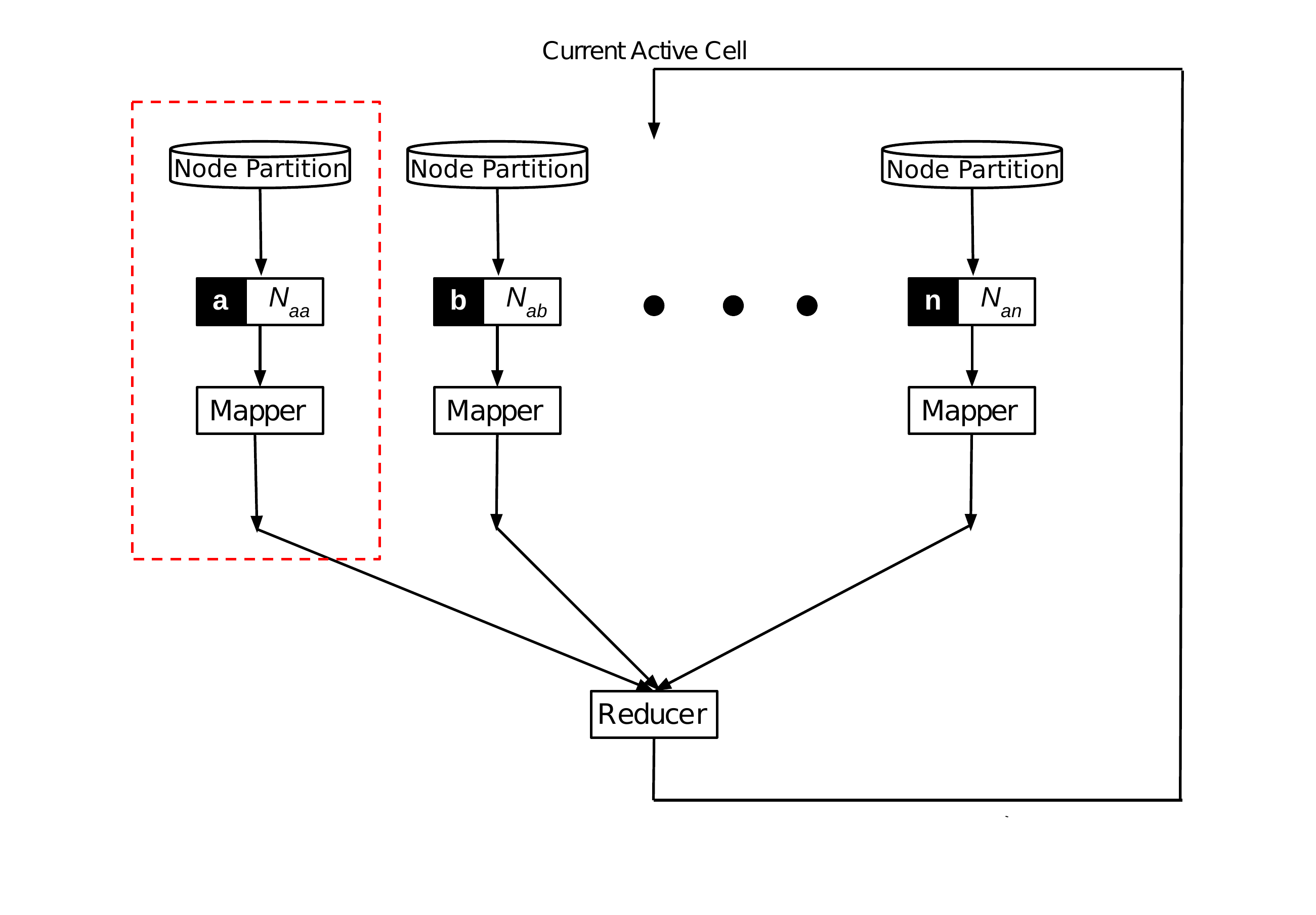}}
	\qquad
	\qquad
	\qquad
	\subfloat[Detailed View of our MapReduce Implementation]{\includegraphics[width=0.8\linewidth]{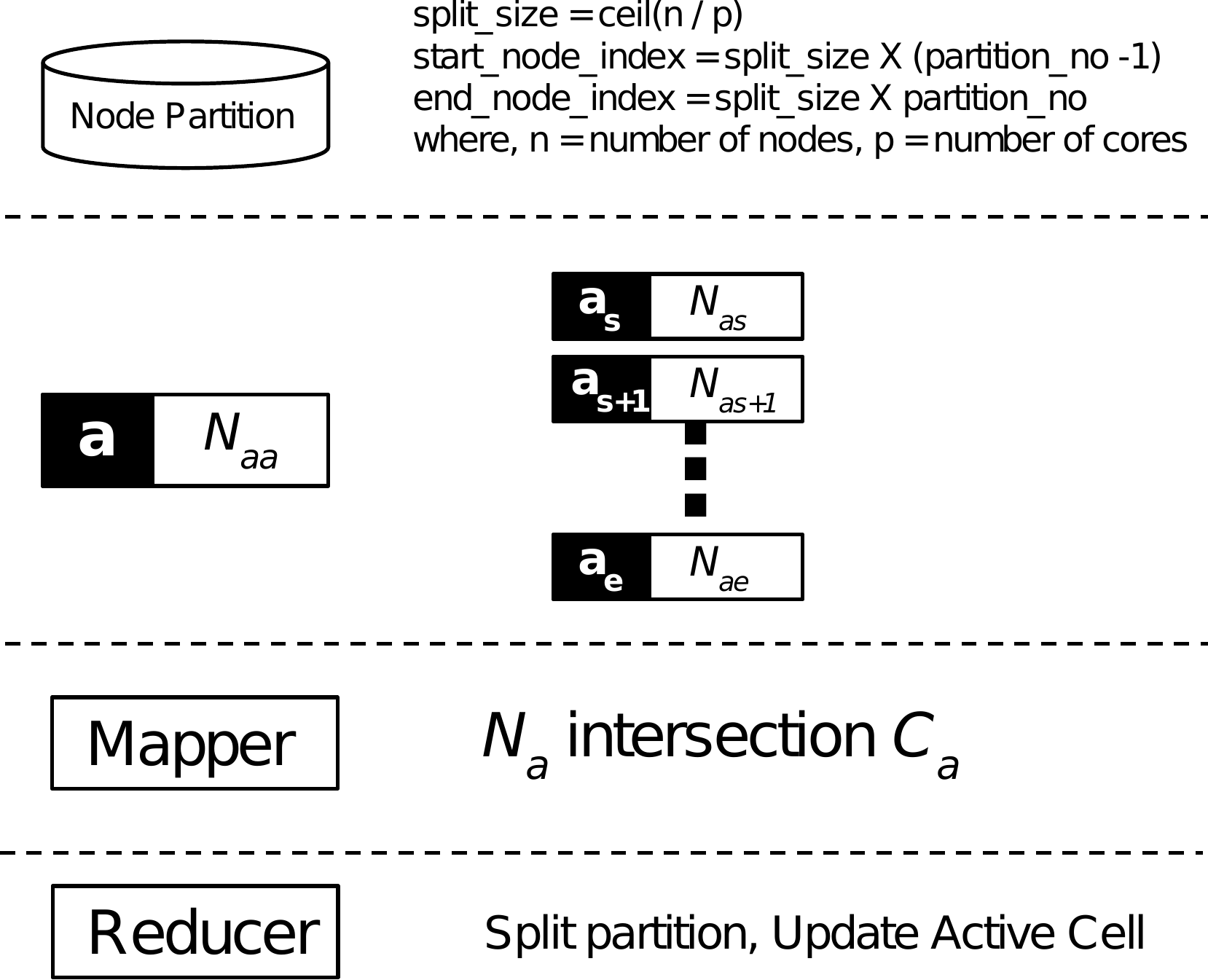}}
	\qquad
	\caption{(a) Conceptual overview of our Lightweight MapReduce implementation. (b) Detailed view of our MapReduce implementation. The data partition number $partition \_ no$ is in the range of $[1,p]$, where $p$ is the number of available cores. The \textit{node partition} splits the input graph into nearly equal sized vertex groups for processing on each of the available cores, we cache the vertex data for each of these groups on the corresponding compute nodes.}
	\label{fig:MR}
\end{figure*}

The proposed Parallel $\epsilon$EP algorithm is \textit{iterative} in nature, which implies that, the output of the current iteration becomes the input for the next one. The number of iterations in the Parallel $\epsilon$EP Algorithm for sparse graphs having a million nodes is in the range of few ten thousands. The existing MapReduce framework implementations such as Hadoop and Disco \cite{hadoop, disco2011} follow the programming model and its architecture from the original MapReduce paradigm \cite{dean2008mapreduce}. Hence, these implementations focus more on data reliability and fault tolerance guarantees in large cluster environments. This reliability and fault tolerance is usually associated with high data copy and job setup overhead costs. Although these features are suited for programs with a single \textsc{map} \& a single \textsc{reduce} operation, they introduce high job setup overhead times across the \textit{iterative} MR steps \cite{twister2010, haloop2010, spark2010}. To circumvent this, we implemented a bare minimum MapReduce framework using open source tools GNU Parallel and rsync \cite{gnu2011, rsync1996}. We used GNU Parallel to trigger parallel map and reduce jobs on the cluster, rsync was used for data copy across the cluster nodes. We were able to achieve job setup overhead time in the range of \textbf{few milliseconds} using the custom framework, as opposed to \textasciitilde$30-45$ seconds for Hadoop on a 10 node cluster isolated over a Gigabit Ethernet switch. Conceptual and detailed overview of our Lightweight MapReduce Framework implementation is depicted in Figure \ref{fig:MR}. We intelligently \textit{sharded} the input graph data across the distributed computing nodes. The \textit{node data partitioning} is performed based on the number of nodes $n$ in the input graph and the number of cores $p$ available for computation; the methodology is depicted in Figure \ref{fig:MR}(b). The \textit{node partition} splits the input graph into nearly equal sized vertex groups for processing on each of the available cores, we cache the vertex data for each
of these groups on the corresponding compute nodes. This is conceptually similar to the user control on \textit{data persistence} and \textit{data partitioning} in the Resilient Distributed Datasets in the \textit{Spark} MapReduce framework \cite{spark2010, rddspark2012}; though our implementation was inspired independently of the Spark framework and realized before that. This helped us achieve locality in reading the input graph. Execution time empirical studies on random power-law graphs for the proposed Algorithm \ref{algo:peep} are presented in Section \ref{sec:scale}.

\section{Experimental Evaluation}
\label{lbl:results}
In this section we present the results of our Fast $\epsilon$EP algorithm. We briefly talk about the datasets used for evaluating our proposed algorithm. We also discuss the evaluation methodology and present our results. Finally, we do the scalability analysis of the proposed Parallel $\epsilon$EP algorithm.

\begin{table*}[htpb]
\small
\vspace{-60pt}
	\parbox{.5\linewidth}{
		\centering
		\caption{\small Facebook Dataset Details}
		\label{tab:facebook}
		\begin{tabular}{|c|c|c|c|} \hline
			Graph&Vertices&Edges&Upto Date\\ \hline
			1&15273&80005&2007-06-22 \\ \hline
			2&31432&218292&2008-04-07 \\ \hline
			3&61096&614796&2009-01-22 \\ \hline
		\end{tabular}
	}
	\parbox{.5\linewidth}{
		\centering
		\caption{\small Flickr Dataset Details}
		\label{tab:flickr}
		\begin{tabular}{|c|c|c|c|} \hline
			Graph&Vertices&Edges&Upto Date\\ \hline
			1&1277145&6042808&2006-12-03 \\ \hline
			2&1856431&10301742&2007-05-19 \\ \hline
		\end{tabular}
	}
\end{table*}

\subsection{Datasets used for Dynamic Analysis}
We have used the \texttt{Facebook} (New Orleans regional network) online social network dataset from \cite{viswanath2009evolution}. The dataset consists of timestamped friendship link formation information between September 26th, 2006 and January 22nd, 2009.  We created three time evolving graph snapshots for the facebook network, the base network consists of all the links formed between September 26th, 2006 and June 22nd 2007. The remaining \textit{two} graphs are created such that, the graph at evolved point of time $t + \delta(t)$ has the graph at time $t$, along with the new vertices and edges that were added to the graph between time $t$ and time $t+\delta(t)$, with $\delta(t)$ being $290$ days. Table \ref{tab:facebook} tabulates the dataset properties.

The second dataset that we used is the \texttt{Flickr} social network dataset from \cite{mislove2008growth}, which consists of a total of 104 days (November 2nd - December 3rd, 2006, and February 3rd - May 18th, 2007) of crawl data. This dataset consists of the timestamped link formation information among nodes. Since the nature of contact links in Flickr are directional in nature, we create an undirected dataset as described next. For each outgoing link  from  user $a\rightarrow b$, if user $b$ reciprocates the link $b\rightarrow a$, we create an undirected edge $(a,b)$. The time of link reciprocation by $b$ is marked as the timestamp of the link formation. Further, we create a time evolving snapshot from this graph. The base graph $G_1$ has data from the first crawl, \textit{i.e.}, between Nov 2nd - Dec 3rd, 2006. The second graph is created in a similar fashion as the \texttt{Facebook} graphs, with $G_2$ being $G_1$ plus the augmented data from the second crawl, \textit{i.e.}, between Feb 3rd - May 18th, 2007. Table \ref{tab:flickr} tabulates the dataset properties.

\subsection{Evaluation Methodology}
We are primarily interested in studying the effect of PA on dynamic social networks and to characterize what role PA plays in the co-evolution of nodes in the networks. Given, a social network graph $G_t$ at time $t$ and its evolved network graph $G_{t+\delta t}$, our algorithm would return an $\epsilon$-equitable partitioning $\pi_t$ for $G_t$ and $\pi_{t+\delta t}$ for $G_{t+\delta t}$. The methodology used to evaluate our proposed $\epsilon$EP algorithm is as follows.
\vspace{-5pt}
\renewcommand{\theenumi}{\roman{enumi}}%
\begin{enumerate}
	\item \textbf{Partition Similarity}: We find the fraction of actors who share the same position across the partitions $\pi_t$ and $\pi_{t+\delta t}$ using Equation \ref{eq:partsim}. The new nodes in $G_{t+\delta t}$, which are not present in $G_t$ are dropped off from $\pi_{t+\delta t}$ before computing the partition similarity score.
	\vspace{-5pt}
	\begin{multline}
		\label{eq:partsim}
		sim(\pi_t, \pi_{t+\delta t}) = \\
		\frac{1}{2} \biggl[\biggl(\frac{N-|\pi_t \cap \pi_{t+\delta t}|}{N-|\pi_t|}\biggr) + \biggl(\frac{N-|\pi_t \cap \pi_{t+\delta t}|}{N-|\pi_{t+\delta t}|}\biggr)\biggr]
	\end{multline}
	where, $N$ is the size of the \textit{discrete} partition of $\pi_t$. The quantity $|\pi_t \cap \pi_{t+\delta t}|$ is the size of the partition obtained by doing \textit{cell-wise} intersection among the cells of $\pi_t$ and $\pi_{t+\delta t}$. In equation \ref{eq:partsim}, if the number of actors who share positions across $\pi_t$ and $\pi_{t+\delta t}$ is large, the value of $|\pi_t \cap \pi_{t+\delta t}|$ will be almost equal to the size of either $\pi_t$ or $\pi_{t+\delta t}$. Hence, the resulting partition similarity score will be close to $1$. On the other hand, if the overlap of actors between $\pi_t$ and $\pi_{t+\delta t}$ is very small, $|\pi_t \cap \pi_{t+\delta t}|$ will be a large number, resulting in a similarity score close to $0$. The terms in the denominator of the equation essentially provide a normalization \textit{w.r.t.} the size of partitions $\pi_t$ and $\pi_{t+\delta t}$. The value of $sim(\pi_t, \pi_{t+\delta t})$ given by Equation \ref{eq:partsim} is always between $[0,1]$. We propose a fast algorithm to compute the partition similarity score in Appendix \ref{app:similarity}.
	\vspace{-5pt}
	\item \textbf{Graph theoretic network centric properties}: Given co-evolving vertex pairs $(a,b)$ which occupy the same position in the partition $\pi_t$, we study the evolution of network centric properties corresponding to the vertex pairs in the time evolved graph $G_{t+\delta t}$. We study the following properties which are widely used for characterizing the graph structure:
	\begin{itemize}
		\item \textit{Betweenness centrality} of a node $v$ is the number of shortest paths across all the node pairs that pass through $v$. This signifies the importance of a node, for routing, information diffusion, etc.
		\item \textit{Degree centrality} of a node is the number of nodes it is connected to in a graph. It quantifies the importance of a node \textit{w.r.t.} the number of connections a node has.
		\item Counting the \textit{number of triangles} a node is part of, is an elementary step required to find the clustering co-efficient of a node in a graph. Clustering co-efficient of a node signifies how strongly-knit a node is with its neighbourhood. There is a scalable algorithm to count the number of triangles in a graph \cite{triangles2011}.
		\item \textit{Shapley value centrality} corresponds to a game theoretic notion of centrality. This models the importance of a node in information diffusion \cite{shapley2013}, it is also efficiently computable for large graphs.  
	\end{itemize}
	We evaluate the co-evolution of nodes in various positional analysis methods using these four \textit{network centric} properties as follows. For each pair of nodes $(a,b)$ which occupy same position in a partition, we compute the difference $(a_t-b_t)$. Where, $a_t$ and $b_t$ correspond to the score of either of these four properties described previously. For the same pair of nodes, we also compute the difference $(a_{t+\delta t}-b_{t+\delta t})$. The scores $a_{t+\delta t}$ and $b_{t+\delta t}$ correspond to the property score at time $t+\delta t$. Finally, we take an absolute value of the difference of these two quantities, \textit{i.e.}, $|(a_t-b_t)-(a_{t+\delta{t}}-b_{t+\delta{t}})|$. A low value of this quantity therefore signifies that for a co-evolving node pair $(a,b)$ at time $t$, the network centric property of node $a$ and $b$ at time $t+\delta t$ have also evolved similarly. Note that we are not partitioning based on the centrality scores, hence our comparisons are across timestamps.
\end{enumerate}

\begin{figure*}[!pb]
\vspace{-255pt}
	\includegraphics[width=\linewidth]{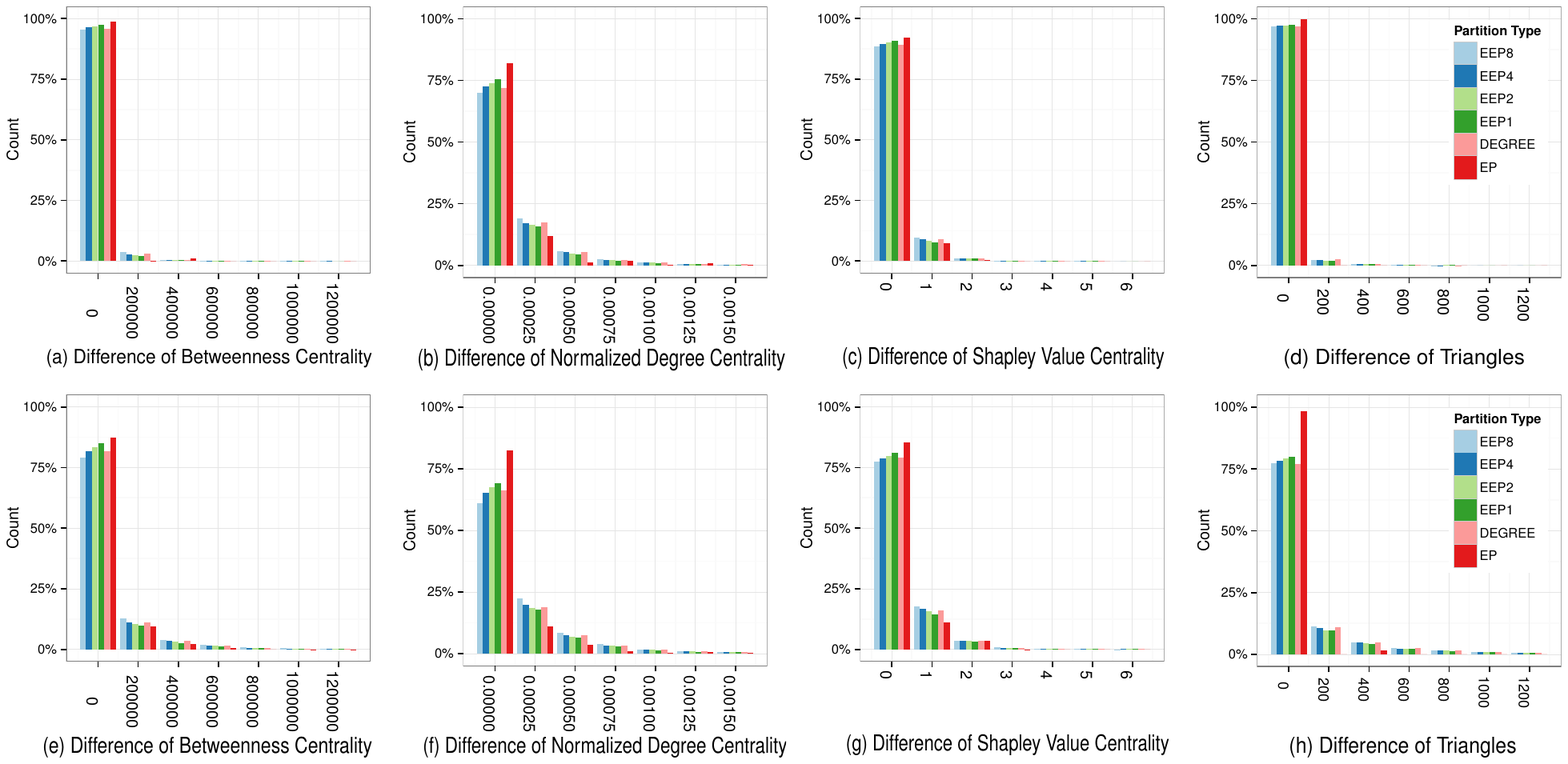}
		\vspace{-225pt}
 	\caption{Co-evolving node pairs for Facebook graphs. Plots (a) to (d) $G_1 \rightarrow G_2$, Plots (e) to (h) $G_1 \rightarrow G_3$.}
 	\label{fig:fbg1g3}
\end{figure*}

\begin{figure*}[!htpb]
	\includegraphics[width=\linewidth]{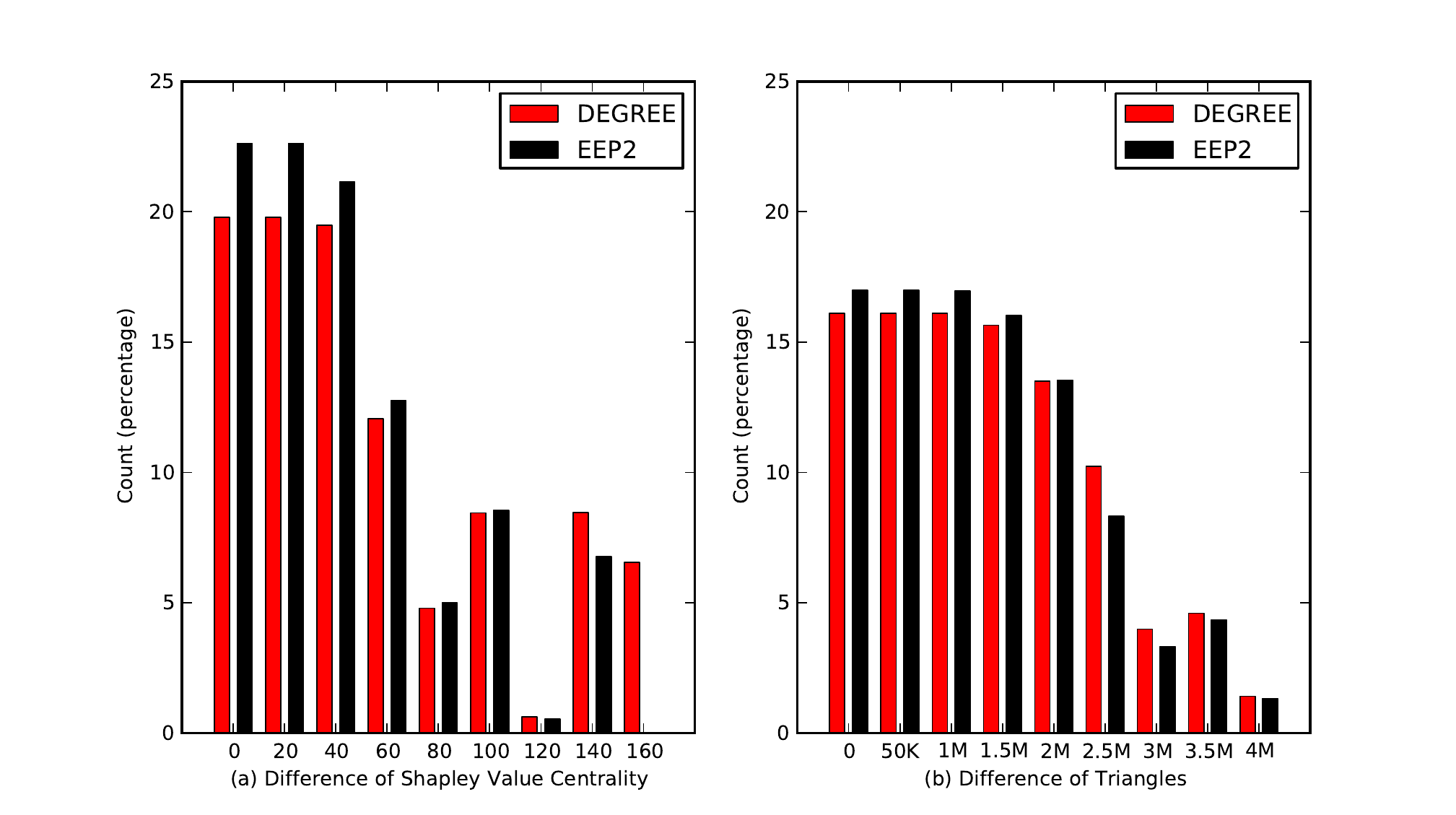}
	\caption{Co-evolving node pairs for Flickr graphs $G_1 \rightarrow G_2$.}
	\label{fig:flickr}
\end{figure*}

\begin{figure*}[!htpb]
	\includegraphics[width=\linewidth]{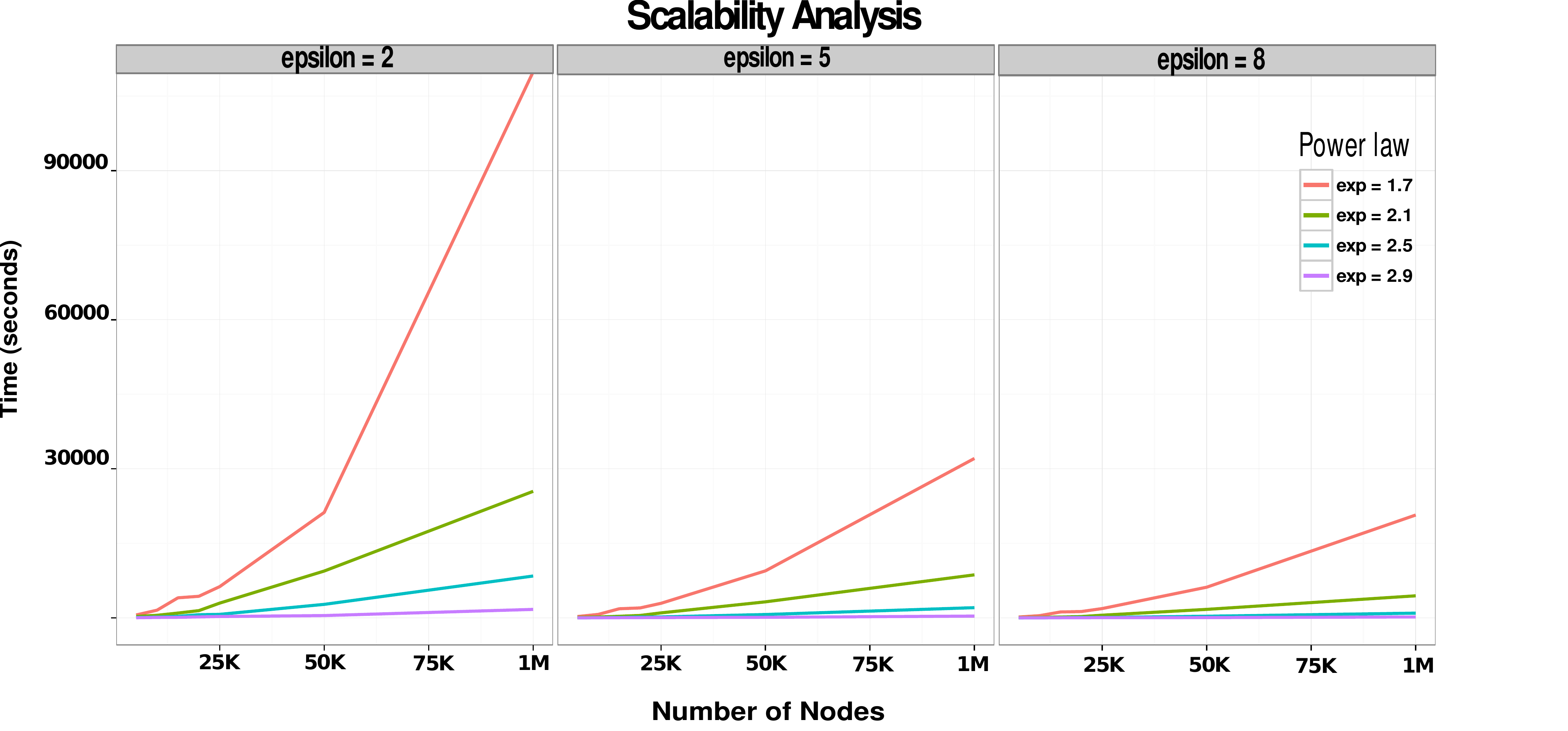}
 	\caption{Scalability Curve for Size of the Input vs Time, for varying Power law exponents and $\epsilon=2,5,8$.}
 	\label{fig:scale}
\end{figure*}

\begin{table*}[htpb]
\begin{centering}
\begin{tabular}{|l|r|r|r|r|r|r|r|r|r|r|}
	\hline
		\textbf{Epsilon ($\epsilon$)} & \textbf{0} & \textbf{1} & \textbf{2} & \textbf{3} & \textbf{4} & \textbf{5} & \textbf{6} & \textbf{7} & \textbf{8} & \textbf{d$^*$}\\ \hline
	\textbf{Graph1 with Graph2} & 59.59 & 66.19 & 76.60 & 83.00 & 86.57 & 89.43 & 91.29 & 92.88 & 94.18 & 86.93\\ \hline
 	\textbf{Graph1 with Graph3} & 54.11 &  57.17 &  69.33 & 76.61 & 80.85 & 84.37 & 86.60 & 88.95 & 90.72 & 79.42\\ \hline
	\textbf{Graph2 with Graph3} & 56.88 & 67.18 & 76.80 & 82.12 & 85.55 & 87.99 & 89.87 & 91.48 & 92.93 & 78.11\\ \hline
\end{tabular} 
\end{centering}
\caption{Percentage of $\epsilon$EP overlap using the Partition Similarity score (Equation \ref{eq:partsim}) for time evolving graphs of the Facebook Network. $\epsilon$ varied from $0$ to $8$, $\epsilon=0$ corresponds to an equitable partition. d$^*$ denotes the partition based on degree.}
\label{tab:fbresult}
\end{table*}

\subsection{Results of Dynamic Analysis}
We present the evaluation of our proposed algorithm using the methodology described in the previous subsection. The results of the \textit{partition similarity} score in percentages are tabulated in Table \ref{tab:fbresult}. We compare our method with \textit{equitable partition} (EP) and the \textit{degree partition}\footnote{\scriptsize Nodes having same degree occupy same position in the partition.} (DP) for the \texttt{Facebook} dataset. We study the evolution of actors from graph $G_1 \rightarrow G_2$, $G_1 \rightarrow G_3$ and $G_2 \rightarrow  G_3$, under these three partitioning methods. The $\epsilon$-equitable partition and the degree partition display a high percentage of overlap among positions than the equitable partition. The poor performance of the EP under the partition similarity score is attributed due the strict definition of equivalence under EP. As an example, consider two nodes $a$ and $b$ occupy same position under EP for graph $G_1$, implies that both have exactly the same degree vector. Suppose, in $G_2$, the number of connections of $b$ remained exactly the same, but node $a$ added one extra link to another position, implies that $a$ and $b$ will now belong to different positions under EP. The $\epsilon$EP consistently performs better than the DP for higher values of $\epsilon$. The higher values of $\epsilon$, correspond to greater bounded relaxation under the definition of $\epsilon$EP. In most of the cases for a given graph, the number of positions under $\epsilon$EP would decrease as we increase the $\epsilon$. Therefore, given two $\epsilon$EPs $\pi_1$ and $\pi_2$, both of them would have relatively less number of positions at higher values of $\epsilon$. This explains the higher partition similarity percentages for $\epsilon$EP. The high values of partition similarity score for degree partition could be attributed due to the nodes in the network which don't evolve with time.

The question on choosing a correct value of $\epsilon$, which corresponds to suitable notion of positions, while satisfying stronger cohesion among actors occupying these positions in dynamic networks is beyond the purview of this paper. Nevertheless results from Table \ref{tab:fbresult} highlight a very important property of $\epsilon$-equitable partition, namely ``\textit{tunability}''.

The study of the various \textit{network centric properties} for co-evolving node pairs of the \texttt{Facebook} and the \texttt{Flickr} datasets, for different positional analysis methods is presented in Figure \ref{fig:fbg1g3} and Figure \ref{fig:flickr} respectively. The \textit{x-axis} corresponds to the bins containing the difference of a network centric property. The \textit{y-axis} corresponds to the frequency of node pairs that belong to a particular bin, as a percentage of the total number of node pairs that occupy the same position in the partition. The results show that equitable partitioning outperforms both the $\epsilon$EP and the DP for each of the network centric properties, which implies that they model positions of co-evolving node pairs pretty well. But the fact that equitable partition leads to trivial partitioning of nodes in a network, makes it the least suitable method for performing PA on real-world networks. Let us consider the example of the equitable partition for the graph $G_3$ from the \texttt{Facebook} dataset which has $61096$ nodes. The EP of $G_3$ has $59474$ cells, out of which $58494$ (\textasciitilde$96\%$) cells are singletons. The co-evolving node pairs under the $\epsilon$EP outperform the DP in most of the cases for the \texttt{Facebook} networks $G_1 \rightarrow G_2$ and $G_1 \rightarrow G_3$, especially for smaller values of $\epsilon$. The $\epsilon$EP with smaller values for $\epsilon$ perform better because of their closeness to the equitable partition. It is worth mentioning here that, the number of positions given by $\epsilon$EP for $\epsilon=1$ for $G_3$ is $8783$. This implies that, $\epsilon$EP guarantee a high degree of confidence on the values of network centric properties of the co-evolving node pairs, along with a partitioning of a reasonable size. The \texttt{Flickr} dataset results in Figure \ref{fig:flickr} also follow a similar trend, the $\epsilon$EP partition performs better than the DP. The percentage counts of both the properties is more spread out across initial few bins for the \texttt{Flickr} dataset. The $\epsilon$EP for $\epsilon=2$ has higher percentage counts for co-evolving node pairs in the bins, which correspond to smaller difference values, whereas, the co-evolving node pairs from the DP have relatively lower percentage counts in the bins closer to a difference of zero and high percentage of nodes towards the tail end of the \textit{x-axis}, especially, for the shapley value centrality, which is not desirable. Also, the degree based partitions give very few positions. Therefore, $\epsilon$EP is a consistent performer, both from the perspective of node co-evolution characteristics and number of positions it gives.

\subsection{Scalability Analysis of the Parallel $\epsilon$EP Algorithm}
\label{sec:scale}
In this section we present empirical studies on the scalability of our proposed parallel algorithm \ref{algo:peep}. The algorithm execution was done on a single machine having eight cores; utilizing all the eight cores for the program. We study the effect of increasing the size of the input, on the running time of the algorithm. We do this analysis on \textit{random power law} graphs by varying the power law exponent $\gamma$ between $1.7 \leq \gamma \leq 2.9$. Figure \ref{fig:scale} shows the various scalability curves. The size of the input varies from $25$ thousand nodes to $1$ million nodes. The running time of the algorithm increases as we decrease the value of $\epsilon$, this is attributed due to the fact that for small values of $\epsilon$, the number of \textsc{splits} which we do (Algorithm \ref{algo:feep}, \textit{line 11}) is quite large, which directly translates to increase in the number of iterations for the algorithm. Also, decreasing the power law exponent $\gamma$, increases the running time of the algorithm. Since, a lower value of $\gamma$ corresponds to \textit{denser} graphs, for dense graphs, the computation of the degree of each vertex to the current active cell $c_a$, therefore, becomes a costly operation. The graph strongly suggests that the algorithm scales almost linearly with increase in the input graph size for the values of $2.1 \leq \gamma \leq 2.9$. It is worth mentioning here that, for most of the real-world graphs, $\gamma$ lies between $2$ and $3$ \cite{powerlaw2009}, with few exemptions. We also performed curve-fitting using polynomial regression to get a complexity bound on the algorithm for $\epsilon=5$. We get a running time bound of $O(n)$, $O(nlogn)$ and $O(n^2)$ for random power law graphs generated using $\gamma=2.9$, $2.5$ and $2.1$ respectively. It is worth noting that, the sum squared residual for $\gamma=2.1$ and the curves $nlogn$ and $n^2$ was quite marginal.
\section{Conclusion and Future Work}
\label{lbl:conclusion}
In this paper we have presented a scalable and distributed $\epsilon$-equitable partition algorithm. To the  best of our knowledge, this is the first attempt at doing positional analysis on a large scale online social network dataset. We have been able to compute $\epsilon$EP for a significantly large component of the \texttt{Flickr} social graph using our Parallel $\epsilon$EP algorithm and its implementation. Further, the results of our algorithm on the \texttt{Facebook} and \texttt{Flickr} datasets show that $\epsilon$EP is a promising tool for analyzing the evolution of nodes in dynamic networks. Empirical scalability studies on random power law graphs show that our algorithm is highly scalable for very large \textit{sparse} graphs. \\
In future, it would be interesting to explore the implied advantage of our Parallel $\epsilon$EP Algorithm to find the \textit{coarsest equitable partition} of very large graphs for an $\epsilon=0$. Finding the equitable partition of a graph forms an important intermediate stage in all the practical graph automorphism finding softwares \cite{nauty2009, saucy2008}. Another possible research direction is to explore algorithms for positional analysis of very large graphs using vertex-centric computation paradigms such as Pregel and GraphChi \cite{pregel2010, graphchi2012}.

\section{Acknowledgements} We thank Inkit Padhi, Intern, RISE Lab, IIT Madras for helping prototype the code for the Fast $\epsilon$EP algorithm. We would also like to thank Srikanth R. Madikeri, Research Scholar, DON Lab, IIT Madras for his valuable inputs during discussions on implementing the Parallel $\epsilon$EP code.

\vspace{2pt}
\bibliographystyle{siam}
\bibliography{sigproc}

\appendix
\section{Partition Similarity Score}
\label{app:similarity}
\subsection{Mathematical Preliminaries}
\label{app:simmaths}
This section briefs out few mathematical preliminaries, which form the basis for our partition similarity score. 

Given, graph G $\equiv \langle$V, E$\rangle$, V is the vertex set, E is the edge set and $\pi = \{C_1, C_2, ...,C_{K}\}$ is a partition of V. We define the following for any two partitions of a graph G $\equiv \langle$V, E$\rangle$:

\renewcommand{\labelenumi}{(\roman{enumi})}

\begin{enumerate}

\item Two \textit{partitions} $\pi_{1} \text{ and } \pi_{2}$ are \textbf{equal}, \textit{iff} they both partition the vertex set V of G exactly in the same way of each other. \\
Example,
\[let, \text{ }
 \pi_1 = \{\{v_1,v_2,v_3,v_4\},\{v_5,v_6\},\{v_7\},\{v_8,v_9,v_{10}\}\}
\]
\[
\pi_2 = \{\{v_6,v_5\},\{v_3,v_2,v_4,v_1\},\{v_9,v_8,v_{10}\},\{v_7\}\}.
\]
\[
 then,\text{ }\pi_1 = \pi_2
\]
\textit{i.e.} the order of cells in partition and the order of vertices inside a cell is not important.

\item We define the \textbf{intersection} of two \textit{partitions} $\pi_{1} \text{ and } \pi_{2}$, as a partition containing the cells obtained from the \textit{set intersection} operator applied \textit{cell-wise} to member cells of $\pi_{\epsilon_1} \text{ and } \pi_{\epsilon_2}$ (discarding the \textit{empty} sets).\\
Example,
\[let, \text{ }
 \pi_1 = \{\{v_1,v_2,v_3\},\{v_4,v_5\},\{v_6,v_7,v_8\}\} \text{ and}
\]
\[
 \pi_2 = \{\{v_1,v_2\},\{v_3,v_4,v_5\},\{v_6,v_7\},\{v_8\}\}.
\]
\[
 then,\text{ }\pi_1 \cap \pi_2 = \{\{v_1,v_2\},\{v_3\},\{v_4,v_5\},\{v_6,v_7\},\{v_8\}\}
\]

\item Two \textit{partitions} $\pi_{1} \text{ and } \pi_{2}$ are \textbf{dissimilar}, \textit{iff} their intersection leads to a \textit{discrete} partition. A discrete partition is a one with only \textit{singleton} cells. \\
Example,
\[let, \text{ }
 \pi_1 = \{\{v_1,v_2,v_3\},\{v_4,v_5\}\} \text{ and}
\]
\[
 \pi_2 = \{\{v_1,v_4\},\{v_3,v_5\},\{v_2\}\}.
\]
\[
 then,\text{ }\pi_1 \cap \pi_2 = \{\{v_1\},\{v_2\},\{v_3\},\{v_4\},\{v_5\}\}
\]
Here, $\pi_1 \cap \pi_2$ gives a \textit{discrete} partition.
\end{enumerate}

\subsection{Simplified Representation of the Partition Similarity Score}

Equation \ref{eq:partsim} can be represented in a simplified form as follows:

\begin{multline}
\label{eq:simDerive}
sim(\pi_1, \pi_2) \\
\begin{aligned}
	&= \frac{1}{2} \biggl[\biggl(\frac{N-|\pi_1 \cap \pi_2|}{N-|\pi_1|}\biggr) + \biggl(\frac{N-|\pi_1 \cap \pi_2|}{N-|\pi_2|}\biggr)\biggr] \\
	&= \frac{1}{2} \biggl(1 - \frac{|\pi_1 \cap \pi_2|}{N} \biggr) \biggl[\biggl( \frac{1}{1 - \frac{|\pi_1|}{N}}\biggr) + \biggl( \frac{1}{1 - \frac{|\pi_2|}{N}}\biggr) \biggr] \\
	&= \frac{1}{2}C(\pi_1 \cap \pi_2) \biggl[ \frac{1}{C(\pi_1)} + \frac{1}{C(\pi_2)} \biggr] \\
	&= \frac{C(\pi_1 \cap \pi_2)}{H(C(\pi_1),C(\pi_2))}
\end{aligned}
\end{multline}
Where, 

\hspace{12pt}$C(\pi) = \biggl( 1 - \frac{|\pi|}{N} \biggr)$,

\hspace{12pt}$|\pi| = \text{cardinality of }\pi$,

\hspace{12pt}$N = \text{cardinality of the \textit{discrete} partition of } \pi$,

\hspace{12pt}$\mathit{H}(x,y) = \text{\textit{harmonic mean} of $x$ and $y$}$. \\

The authors in \cite{partitions2006comparison} survey and compare several notions of distance indices between partitions on the same set, which are available in the literature.

\subsection{MapReduce Algorithm to Compute the Partition Similarity Score}
\label{app:simalgo}
The partition similarity score of Equation \ref{eq:partsim} requires the cardinality of the intersection set of the two partitions $\pi_1$ and $\pi_2$. Finding the intersection of two partitions as per the definition of \textit{intersection} from $(ii)$, Appendix \ref{app:simmaths} is $O(n^2)$ operation, where $n$ being the total number of vertices in the partition. Computing this for very large graphs becomes intractable. To counter this problem, we provide an algorithm based on the MapReduce paradigm \cite{dean2008mapreduce} to compute the size of the \textit{intersection set} of $\pi_1$ and $\pi_2$ (\textit{i.e.}, $|\pi_1 \cap \pi_2|$). The algorithm is presented in Algorithm box \ref{algo:partsim}. The algorithm initializes by \textit{enumerating} the indices of $\pi_2$ for each cell index of partition $\pi_1$. For each key from this tuple, the \textsc{map} operation checks if these two cells \textit{intersect} or not. The \textsc{map} emits a \textit{value} of $1$ corresponding to a constant \textit{key}. The \textsc{reduce} operations computes the \textit{sum} of these individual $1s$. This \textit{sum} corresponds to $|\pi_1 \cap \pi_2|$, which is used to compute the partition similarity score from Equation \ref{eq:partsim}.

\textbf{A note on Algorithm \ref{algo:partsim}:} The \textsc{initialize} method of Algorithm \ref{algo:partsim} (\textit{line} 3), primarily involves replicating/enumerating the cell indices of $\pi_2$ to all the cell indices of $\pi_1$. Since \textit{cross} operations are computationally very costly, the tractability of the Algorithm is inherently dependant on ability to generate the \textit{cross product} set of the cell index tuples of the two input partitions. 

\begin{algorithm}[htpb]
	\caption{MapReduce Partitions Intersection Set Cardinality}
	\label{algo:partsim}
	\textbf{Input:} Partitions $\pi_1$ and $\pi_2$. Let $\pi_1$=$\{cell_1$:$[v_1,v_2], cell_2$:$[v_3,v_4,v_5],...,cell_K$:$[v_{n-2},v_{n-1},v_n]\}$ \textbf{and} $\pi_2$=$\{cell_1$:$[v_1,v_2,v_3], cell_2$:$[v_4,v_5],...,cell_L$:$[v_{n-1},v_{n}]\}$\\
	\textbf{Output:} Partitions intersection set cardinality  $|\pi_1 \cap \pi_2|$\\
	\hrulefill
	\begin{algorithmic}[1]
		\State \textbf{class} \textsc{Mapper}
		\Statet \textbf{tupleList} \textit{enumCells}=[ ]
		\Statex
		\Statet  \textbf{method} \textsc{initialize}()
		\Statett \textbf{for each} \textit{cell\_index} \textbf{$i$} of $\pi_1$ (\textit{i.e.} $1\to K$) \textit{enumerate} the \textit{cell\_index} $j$ of $\pi_2$ (\textit{i.e.} $1\to L$) \textbf{do}
		\Statett\hspace{8pt}\textbf{add} $tuple(i,j)$ them to $enumCells$
		\Statex
		\Statet  \textbf{method} \textsc{map}(id t, tuple $(i,j)$)
		\Statett  $intersect\leftarrow\{\pi_1(i)\cap{\pi_2(j)\}}$ \Comment Appendix \ref{app:simmaths}, $(ii)$
		\Statett \textbf{if} $intersect\neq \phi$ \textbf{then}
		\Statett\hspace{8pt} \textsc{emit}(id \textit{intersect}, $1$) \Comment If the two cells have a overlap, emit \textit{value} $1$ corresponding to a constant \textit{key} ``\textit{intersect}''
	\end{algorithmic}
	\hrulefill
	\begin{algorithmic}[1]
		\State \textbf{class} \textsc{Reducer}
		\Statet \textbf{method} \textsc{reduce}(id $key$, values)
		\Statett $sum=0$
		\Statett \textbf{for} \textit{value} in values 
		\Statett\hspace{8pt} $sum=sum+value$
		\Statett \textsc{emit}(\textit{key}, $sum$) \Comment $|\pi_1 \cap \pi_2|$
	\end{algorithmic}
\end{algorithm}

\end{document}